\documentclass[twocolumn,twocolappendix,appendixfloats]{aastex6}

\usepackage{subfigure}
\usepackage{threeparttable}
\usepackage{multirow}
\usepackage{remreset}
\usepackage{CJKutf8}

\collaborationName{Friends of AASTeX}
\shorttitle{A kinematic shift of the C\,{\footnotesize IV} BAL}
\shortauthors{Lu \& Lin}
\begin{document}


\title{A kinematic shift of the C\,{\footnotesize IV} Broad Absorption Line in Quasar SDSS J120819.29+035559.4}


\author{
\begin{CJK*}{UTF8}{gbsn}
Wei-Jian Lu (陆伟坚)\altaffilmark{1} and Ying-Ru Lin (林樱如)\altaffilmark{2}
\end{CJK*}
}
\affil{School of Information Engineering, Baise University, Baise 533000, China}

\altaffiltext{1}{E-mail: william\_lo@qq.com (W-J L)}
\altaffiltext{2}{E-mail: yingru\_lin@qq.com (Y-R L)}



\begin{abstract}
We report the kinematic shift of the C\,{\footnotesize IV} broad absorption line (BAL) in quasar SDSS J120819.29+035559.4 (hereafter J1208+0355). This quasar shows two BAL systems, {including a blue component of system A at $\sim$--23500$\,\rm km\,s^{-1}$ that shows a kinematic shift of --1166$\pm$65 $\,\rm km\,s^{-1}$, and a red component of system B at $\sim$--7000$\,\rm km\,s^{-1}$ that can be decomposed into several narrow absorption lines (NALs). }First, we confirm that the most likely cause for the equivalent width variations of the absorption lines {(at least for system B)} in J1208+0355 is the ionization change scenario as a response to the changes in the ionization continuum according to the following observational factors: (1) coordinated  multiple absorption lines strengthening; (2) the continuum flux shows an obvious weakening. Second, we find line-locking phenomena from the blended NALs within system B, indicating that these outflow clouds are driven by a radiative force caused by resonance lines. The above two research aspects convincingly reveal that the BAL outflows of J1208+0355 are affected by the background radiation energy. Therefore, we infer that the kinematic shift shown in system A may be produced by actual line-of-sight acceleration of the outflow clouds, which is driven by radiation pressure from the background light source.
\end{abstract}

\keywords{Quasars (1319); Broad-absorption line quasar (183); Quasar absorption
line spectroscopy (1317)}



\section{Introduction} \label{sec:intro}
Intrinsic absorption lines of quasars, including broad absorption lines (BALs, with absorption widths of $\textgreater$2000\,$\rm km\,s^{-1}$; e.g., \citealp{Weymann1991}), mini-BALs (loosely defined to have absorption widths from 2000\,$\rm km\,s^{-1}$ to several hundred \,$\rm km\,s^{-1}$; e.g., \citealp{Hamann2004}), and narrow absorption lines (NALs, with absorption widths of a few hundred\,$\rm km\,s^{-1}$), can show variation in their equivalent widths (EWs) across a rest-frame time from several days to several years (e.g., \citealp{He2017, Chen2018b,Hemler2019}, and references therein). However, the signatures of kinematic changes of intrinsic absorption lines have been reported in only a few cases (e.g., \citealp{Vilkoviskij2001,Rupke2002,Gabel2003,Hall2007,Joshi2014,Joshi2019,Grier2016}), although they are key physical properties of the outflow winds.

\begin{figure*}
\includegraphics[width=2\columnwidth]{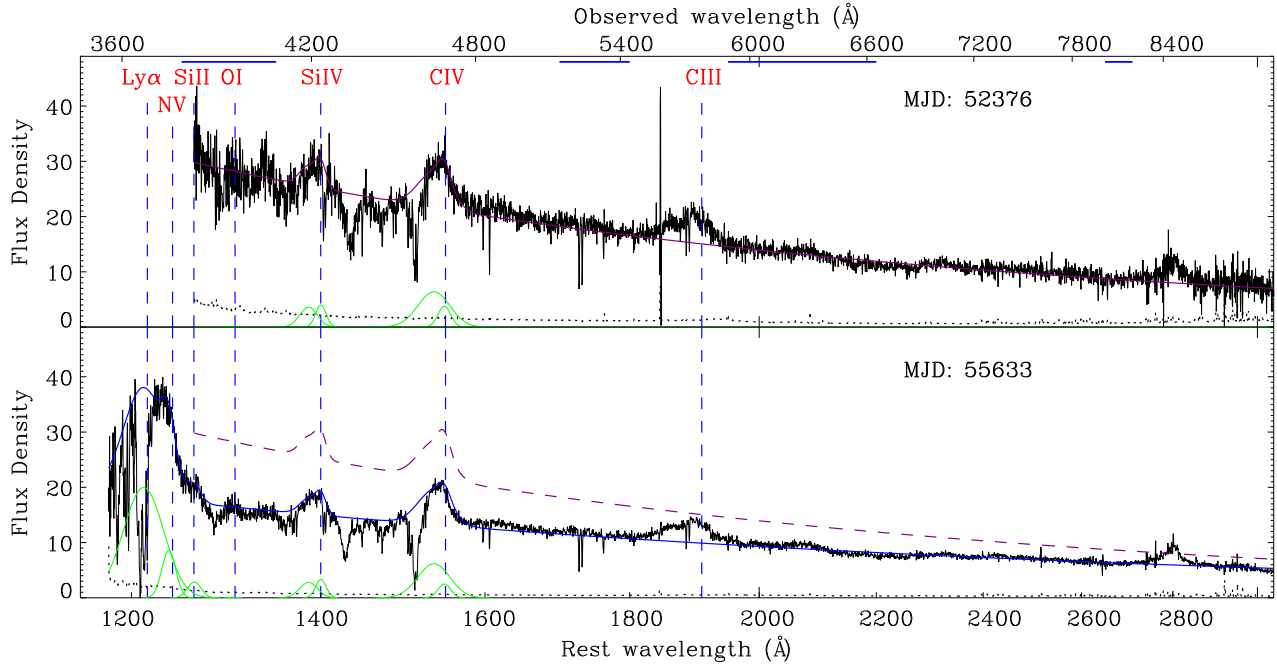}
\caption{Final pseudo-continuum fits (the purple and blue solid curves) for the two spectra of quasar J1208+0355. The unit of flux density is $\rm 10^{-17}~erg~s^{-1}~cm^{-2}$. The main emission lines are marked with blue vertical dashed lines. The blue horizontal bars on the top of the upper panel are the relatively line-free regions used for fitting the power-law continua. {The purple dashed line in the bottom panel is the final pseudo-continuum fit for the MJD
52,376 spectrum.} The dotted lines at the bottom of each panel are the formal 1$\sigma$ errors. Gaussian profiles that in green at the bottom of each panel are the emission line fits. The transverse axes are logarithmic.}  \label{fig.1}
\end{figure*}
\begin{figure*}
\includegraphics[width=2\columnwidth]{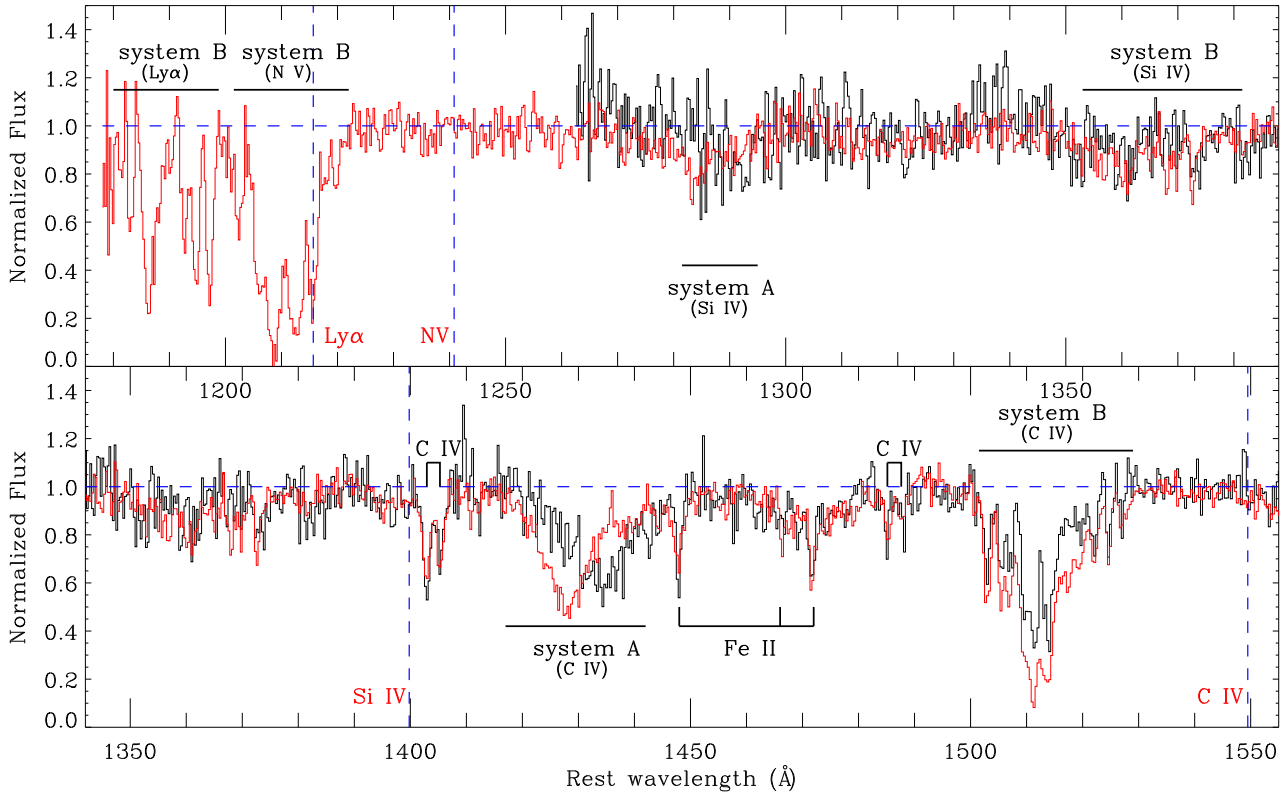}
\caption{Normalized spectra of J1151+0204. The black and red lines, respectively, represent the MJD 52,376 and 55,633 spectra. The black horizontal bars mark the absorption line  systems. The main emission lines are marked with blue vertical dashed lines.} \label{fig.2}
\end{figure*}
\begin{figure}
\includegraphics[width=\columnwidth]{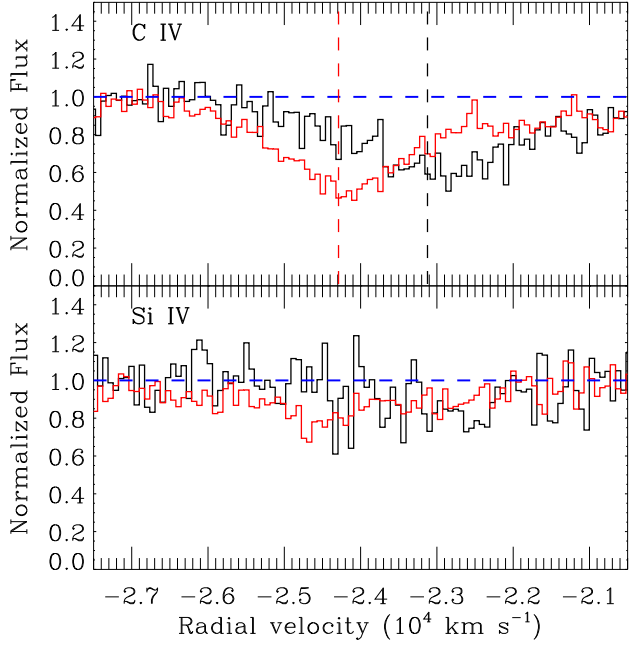}
\caption{Normalized spectra of J1151+0204 showing the C\,{\footnotesize IV} and Si\,{\footnotesize IV} BALs of system A. The black and red lines represent the MJD 52,376 and 55,633 spectra. The black and red horizontal vertical dashed lines mark the line centers of the two-epoch BALs.} \label{fig.3}
\end{figure}
\begin{figure}
\includegraphics[width=\columnwidth]{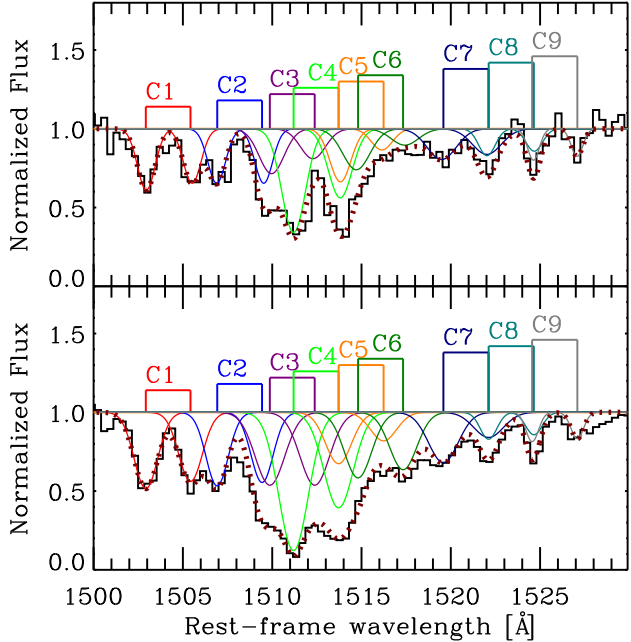}
\caption{Identified NAL doubles within the C\,{\footnotesize IV} BAL of system B.  The top and bottom spectra are respectively snippets from the MJD 52,376 and 55,633 normalized spectra. The brown dotted lines represent the total fit models.} \label{fig.4}
\end{figure}
\begin{figure}
\includegraphics[width=\columnwidth]{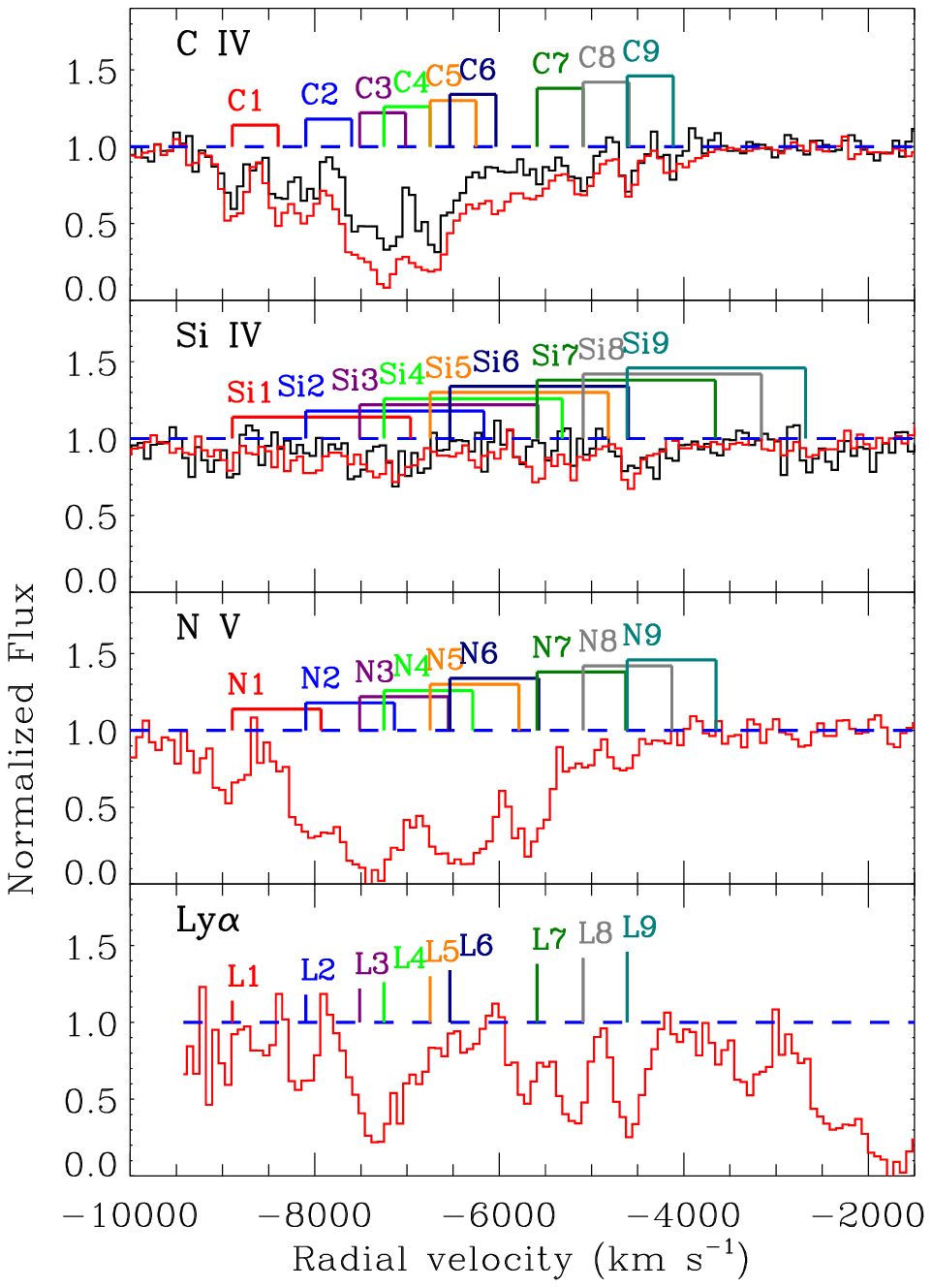}
\caption{Portions of the normalized spectra of J1151+0204, showing the NAL systems identified within system B in different ions. The black and red lines represent the normalized spectra from observations on MJD 52,376 and 55,633, respectively.} \label{fig.5}
\end{figure}

For NALs, \citet{Gabel2003} have firstly detected a synchronous deceleration for C\,{\footnotesize IV}, Si\,{\footnotesize IV}, and N\,{\footnotesize V} NALs, while no change in their absorption profiles has been found, based on the analyses on the Seyfert galaxy NGC 3783 (see also \citealp{Scott2014}). Recently, \citet{Misawa2019} have reported the monitoring of intrinsic NALs in six quasars across timescales of 2.8--5.5 yr (in quasar rest frame) but failed to find an obvious shift in their velocities. For BALs, \citet{Vilkoviskij2001}, \citet{Rupke2002} and \citet{Hall2007} have respectively reported positive kinematic shifts of BALs in individual objects. After that, deceleration in C\,{\footnotesize IV} BALs over rest-frame timescales of 3.11 and 2.34 yr have also been found, from two X-ray-bright quasars \citep{Joshi2014}. The first systematic search for BAL kinematic shifts has been performed based on three-epoch observations of 140 BAL quasars from the Sloan Digital Sky Survey (SDSS; \citealp{York2000}) by \citet{Grier2016}. Three cases out of their quasar sample show a velocity shift in C\,{\footnotesize IV} BALs, which reveals that the BAL velocity shift is not widespread. Moreover, these three cases exhibit an unstable magnitude in the velocity shifts along with time. More recently, \citet{Joshi2019} have reported the deceleration of both C\,{\footnotesize IV} and Si\,{\footnotesize IV} BALs in quasar SDSS J092345+512710.
 
In this work, we report the discovery of the kinematic shift of a BAL system in the two-epoch spectra of quasar SDSS J120819.29+035559.4 (hereafter J1208+0355, $z_{\rm em}$=2.023; \citealp{Paris2018}, ). This source has been studied in previous systematic studies of EW variations in both its BALs \citep{He2017} and NALs \citep{Chen2015a}. Here we will further to confirm the physical mechanism of the EW variation in its BALs and NALs, study the line-locking phenomena of the blended NALs within one of its BAL systems, and study the velocity shift phenomenon shown in another BAL system. Section \ref{sec.2} is the spectral analysis for the two-epoch observations of the source. Section \ref{sec.3} presents discussions on the variation mechanism, line-locking phenomenon, and the kinematic shift. Finally, we draw a conclusion in Section \ref{sec.4}. This paper adopts a $\Lambda$CDM cosmology with parameters $H_0=70\,\rm km\,s^{-1}\,Mpc^{-1}$, $\Omega_{\rm M}=0.3$, and $\Omega_{\Lambda}=0.7$.

\begin{table*}[h]
    \centering
\caption{Measurements of absorption lines \label{tab.1}}
\begin{tabular}{cccccccccc} 
\hline 
\hline 
Species & $z_{\rm abs}$ & Velocity$^{\rm a}$ & \multicolumn2c{MJD:52,376} && \multicolumn2c{MJD:55,633} & Fractional  &Note\\
\cline{4-5}\cline{7-8}
 & & &EW &FWHM$^{\rm b}$ &&EW &FWHM$^{\rm b}$ &EW &    \\
 & &($\rm km~s^{-1}$) & (\AA) & ($\rm km~s^{-1}$) && (\AA) & ($\rm km~s^{-1}$) & Variation & \\
\hline
C\,{\footnotesize IV}$\lambda$1549	&	1.9346	&	$	 	-8893 		$	&	$	0.50 	\pm	0.10 	$	&	233 	&	&	$	0.83 	\pm	0.05 	$	&	311 	&	$	0.50 	\pm	0.196 	$	&	C1	\\
C\,{\footnotesize IV}$\lambda$1551	&	...	&	$	...			$	&	$	0.44 	\pm	0.11 	$	&	233 	&	&	$	0.75 	\pm	0.05 	$	&	311 	&	$	0.52 	\pm	0.241 	$	&	...	\\
C\,{\footnotesize IV}$\lambda$1549	&	1.9424	&	$	 	-8099 		$	&	$	0.44 	\pm	0.11 	$	&	232 	&	&	$	0.75 	\pm	0.05 	$	&	291 	&	$	0.52 	\pm	0.241 	$	&	C2	\\
C\,{\footnotesize IV}$\lambda$1551	&	...	&	$	...	 		$	&	$	0.44 	\pm	0.10 	$	&	232 	&	&	$	0.71 	\pm	0.04 	$	&	291 	&	$	0.47 	\pm	0.221 	$	&	...	\\
C\,{\footnotesize IV}$\lambda$1549	&	1.9482	&	$	 	-7513 		$	&	$	0.51 	\pm	0.14 	$	&	325 	&	&	$	0.98 	\pm	0.04 	$	&	387 	&	$	0.63 	\pm	0.250 	$	&	C3	\\
C\,{\footnotesize IV}$\lambda$1551	&	...	&	$	...	 		$	&	$	0.34 	\pm	0.20 	$	&	325 	&	&	$	0.98 	\pm	0.05 	$	&	387 	&	$	0.97 	\pm	0.452 	$	&	...	\\
C\,{\footnotesize IV}$\lambda$1549	&	1.9508	&	$	 	-7250 		$	&	$	1.12 	\pm	0.06 	$	&	309 	&	&	$	1.86 	\pm	0.02 	$	&	386 	&	$	0.50 	\pm	0.051 	$	&	C4	\\
C\,{\footnotesize IV}$\lambda$1551	&	...	&	$	...	 		$	&	$	0.75 	\pm	0.08 	$	&	309 	&	&	$	1.29 	\pm	0.03 	$	&	386 	&	$	0.53 	\pm	0.102 	$	&	...	\\
C\,{\footnotesize IV}$\lambda$1549	&	1.9557	&	$	 	-6751 		$	&	$	0.51 	\pm	0.09 	$	&	274	&	&	$	0.61 	\pm	0.06 	$	&	339 	&	$	0.18 	\pm	0.200 	$	&	C5	\\
C\,{\footnotesize IV}$\lambda$1551	&	 ... 	&	$	...	 		$	&	$	0.20 	\pm	0.27 	$	&	274	&	&	$	0.34 	\pm	0.13 	$	&	339 	&	$	0.52 	\pm	1.309 	$	&	...	\\
C\,{\footnotesize IV}$\lambda$1549	&	1.9578	&	$	 	-6535 		$	&	$	0.51 	\pm	0.15 	$	&	355	&	&	$	0.81 	\pm	0.05 	$	&	355 	&	$	0.45 	\pm	0.285 	$	&	C6	\\
C\,{\footnotesize IV}$\lambda$1551	&	...	&	$	...	 		$	&	$	0.20 	\pm	0.42 	$	&	355	&	&	$	0.71 	\pm	0.06 	$	&	355 	&	$	1.12 	\pm	1.442 	$	&	...	\\
C\,{\footnotesize IV}$\lambda$1549	&	1.9671	&	$	 	-5592 		$	&	$	0.40 	\pm	0.23 	$	&	377	&	&	$	0.67 	\pm	0.07 	$	&	384 	&	$	0.50 	\pm	0.547 	$	&	C7	\\
C\,{\footnotesize IV}$\lambda$1551	&	...	&	$	...	 		$	&	$	0.34 	\pm	0.27 	$	&	377	&	&	$	0.34 	\pm	0.15 	$	&	384 	&	$	0.00 	\pm	0.908 	$	&	...	\\
C\,{\footnotesize IV}$\lambda$1549	&	1.9721	&	$	 	-5093 		$	&	$	0.20 	\pm	0.18 	$	&	215	&	&	$	0.20 	\pm	0.10 	$	&	215 	&	$	0.00 	\pm	1.030 	$	&	C8	\\
C\,{\footnotesize IV}$\lambda$1551	&	...	&	$	...			$	&	$	0.17 	\pm	0.22 	$	&	215	&	&	$	0.17 	\pm	0.12 	$	&	215 	&	$	0.00 	\pm	1.474 	$	&	...	\\
C\,{\footnotesize IV}$\lambda$1549	&	1.9768 	&	$	 	-4612 		$	&	$	0.17 	\pm	0.10 	$	&	153	&	&	$	0.17 	\pm	0.07 	$	&	164 	&	$	0.00 	\pm	0.718 	$	&	C9	\\
C\,{\footnotesize IV}$\lambda$1551	&	...	&	$	...			$	&	$	0.15 	\pm	0.12 	$	&	153	&	&	$	0.15 	\pm	0.10 	$	&	164 	&	$	0.00 	\pm	1.041 	$	&	...	\\
C\,{\footnotesize IV} BAL 	&	...	&	$	-25466 	\sim	-21515 	$	&	$	4.96 	\pm	0.32 	$	&	3952 	&	&	$	...	 	 	$	&	...	&	$	...			$	&	system A	\\
C\,{\footnotesize IV} BAL 	&	...	&	$	-26724 	\sim	-22758 	$	&	$	...			$	&	...	&	&	$	5.20 	\pm	0.18 	$	&	3966 	&	$	0.05 	\pm	0.073 	$	&	system A	\\
C\,{\footnotesize IV} BAL 	&	...	&	$	-9426 	\sim	-3989 	$	&	$	6.52 	\pm	0.31 	$	&	5437 	&	&	$	10.47 	\pm	0.16 	$	&	5437 	&	$	0.47 	\pm	0.047 	$	&	system B	\\
Si\,{\footnotesize IV}BAL 	&	...	&	$	-10191 	\sim	-3945 	$	&	$	1.97 	\pm	0.47 	$	&	6246 	&	&	$	3.27 	\pm	0.27 	$	&	6246 	&	$	0.50 	\pm	0.235 	$	&	system B	\\
N\,{\footnotesize V} BAL  	&	...	&	$	-9648 	\sim	-4579 	$	&	$	...			$	&	... 	&	&	$	10.09 	\pm	0.26 	$	&	5069 	&	$	...			$	&	system B	\\
ly$\alpha$ BAL  	&	...	&	$	-8925 	\sim	-4189 	$	&	$	...			$	&	... 	&	&	$	5.54 	\pm	0.41 	$	&	4736 	&	$	...			$	&	system B	\\
\hline 
\end{tabular}
\begin{tablenotes}
\footnotesize
\item$^{\rm a}$Velocity range of the BAL troughs with respect to the emission rest frame.
\item$^{\rm b}$Total width calculated from edge-to-edge of the BAL trough.
\end{tablenotes}
\end{table*}

\section{SPECTROSCOPIC ANALYSIS and results} \label{sec.2} 
We downloaded the two-epoch spectra of J1208+0355 from the SDSS for the MJD 52,376 spectrum, and from the Baryon Oscillation Spectroscopic Survey (BOSS; \citealp{Dawson2013}) for the MJD 55,633 spectrum. The SDSS spectrum covers a wavelength range from $\thicksim$3800 to 9200 \,\AA~and has a resolution within$\thicksim$1850--2200, while the BOSS covers wavelengths between $\thicksim$3600 and 10000\,\AA~and has a resolution ranging from $\thicksim$1300 to 3000. The median signal-to-noise ratios (S/Ns) are 13.90 and 21.08 per pixel for the MJD 52,376 and 55,633 spectra of J1208+0355, respectively.

The procedures for spectroscopic analysis are the same as those in our previous works (e.g., \citealp{Lu2018complex2,Lu2018saturation}). In short, the power-law continua were acquired from the iterative fitting of several wavelength regions (1250--1350, 1700--1800, 1950--2200, and 2650--2710\,\AA~in the rest frame), which were defined by \citet{Gibson2009a}. Then we combined each power-law continuum fit with the Gaussian fits of the emission lines, thus obtaining the final pseudo-continuum (Figure \ref{fig.1}). Finally, we normalized the original spectra by using these pseudo-continua (Figure \ref{fig.2}).

Shown in Figure \ref{fig.2} are the normalized spectra of J1208+0355, from which two BAL systems were detected, including a blue component that shows kinematic shift and a red component that can be decomposed into several NALs (hereafter systems A and B, respectively). The velocities, EWs, FWHMs, and fractional EW variations of the BALs of these two systems are all listed in Table \ref{tab.1}. The methods of the calculations for the line EWs, as well as their corresponding errors, are the same as those in Lu \& Lin (\citeyear{Lu2018complex1}, see their equations (2) and (3)), and those for the fractional EW variation are the same as those in Lu et al. (\citeyear{LLQ2018}, see their equations (2) and (3)).

In system A, we found a velocity shift between the two-epoch observations of the C\,{\footnotesize IV} BAL. Although the Si\,{\footnotesize IV} BAL of system A also shows a faint sign of kinematic shift, it suffers from a relatively low S/N and is too weak to be identified. We consider the position that averages the EW of a BAL as its line center. To estimate the line center more correctly, we generated 10,000 mock spectra by adding Gaussian noise to the original spectrum according to the corresponding flux density errors, and then measured the line center for these mock spectra using the same procedure. Finally, we estimated the line center and the corresponding measurement uncertainties from the median value and the standard deviation of the 10,000 trials, respectively. We got the line centers of $-23123\pm$100\,$\rm km\,s^{-1}$ and $-24289\pm$47\,$\rm km\,s^{-1}$ for the MJD 52,376 and 55,633 C\,{\footnotesize IV} BALs of system A, respectively, and their velocity shift is --1166$\pm$65\,$\rm km\,s^{-1}$, corresponding to an acceleration rate of 1.253$\pm$0.070\,$\rm cm~s^{-2}$ (Figure \ref{fig.3}). 

In the C\,{\footnotesize IV} BAL of system B, there are at least nine pairs of C\,{\footnotesize IV} NAL doubles that could be well fitted by Gaussian functions (Figure \ref{fig.4}). The line center positions of the corresponding NAL components in Si\,{\footnotesize IV}, N\,{\footnotesize V}, and Ly$\alpha$ ions were also marked (Figure \ref{fig.5}), which were determined by anchoring their blue members to those of C\,{\footnotesize IV} lines. The parameters of these NALs that were measured from the Gaussian functions are also presented in Table \ref{tab.1}. 
\begin{table}
    \centering
\caption{Velocity Splitting among Distinct NAL Systems\label{tab.2}}
\begin{tabular}{lcrcccrc} 
\hline 
\hline 
Doublet &\multicolumn3c{$\rm Splitting^a$} & Components & \multicolumn3c{$\rm Separation^b$} \\
             &\multicolumn3c{(\,$\rm km\,s^{-1}$)} & &\multicolumn3c{(\,$\rm km\,s^{-1}$)} \\
\hline
C\,{\footnotesize IV}&&499&&4--5&&500&\\
                                         &&&&7--8&&499&\\ 
                                         &&&&8--9&&479&\\ 
Si\,{\footnotesize IV}&&1933&&3--7&&1923&\\                                                                                
                                         &&&&6--9&&1921&\\ 
N\,{\footnotesize V}&&962&&3--6&&979&\\
                                          &&&&6--7&&943&\\
                                          &&&&7--9&&977&\\
S\,{\footnotesize IV}&&2894&&3--9&&2900&\\
O\,{\footnotesize VI}&&1649&&1--4&&1645&\\
                                          &&&&4--7&&1658&\\
                                          &&&&5--8&&1657&\\
\hline 
\end{tabular}
\begin{tablenotes}
\footnotesize
\item{$^{\rm a}$The laboratory value.}
\item{$^{\rm b}$The measured separation in the spectra.}
\end{tablenotes}
\end{table}

\section{DISCUSSION} \label{sec.3}
\subsection{{Variation Mechanism }} \label{sec.3.1} 
Previous research considers that EW variation of BALs may be caused by two scenarios: the ionization change (IC) scenario and the transverse motion (TM) scenario of the absorption materials. Our viewpoint is that the IC scenario is more likely to be the mechanism that is responsible for the absorption line variation in J1208+0355, due to at least the following several factors. 

First, all the absorption lines of J1208+0355 show coordinated strengthening between the two-epoch observations. This strengthening includes that between the multiple NAL systems (C1$\sim$C9) within system B, and that between the systems A and B. Such a phenomenon seems unrealistic to explained with the TM scenario, because the coordinated variations of so many distinct absorption systems would require highly coordinated motions between many outflow clouds (e.g., \citealp{Misawa2005,Hamann2011}). {We note that in comparison with the C\,{\footnotesize IV} BAL of system B that shows {nearly 50\% fractional EW variation}, the change of the C\,{\footnotesize IV} BAL of system A is much smaller (only about 5\%) and is consistent with no change. One possible explanation could be that the C\,{\footnotesize IV} BAL of system A may suffer from saturation. A saturated BAL might respond to the continuum fluctuations softly or might even not respond under the situation with the IC scenario as its variation mechanism.}

Second, the continuum flux of the source shows obvious weakening. Correlations between the variations of the ionizing continuum and those of outflow absorption lines have already been proved (\citealp{Lu2017,Chen2018a,Chen2018b,Chen2019} for NALs; \citealp{LLQ2018,Lu2018saturation,Lu2019individual,Huang2019,Vivek2019} for BALs), which could, at least to some extent, serve as an evidence for supporting the IC scenario as the main origin of absorption line variability. For J1208+0355, along with strengthenings of the absorption lines, the power-law continuum (the purple and blue solid curves in Figure \ref{fig.1}) in {the second epoch is on average 41\% lower than in the first epoch.} Such a situation is compatible with the previous reported anticorrelations. Photoionization simulations show that with an increasing ionization parameter ($U$), the  C\,{\footnotesize IV} and Si\,{\footnotesize IV} EWs would rise first, then will reach a peak, and finally decrease (e.g., \citealp{He2017}). Assuming that the IC scenario is indeed the variation mechanism for the outflow lines of J1208+0355, then according to the photoionization simulation {(see figure 3 of \citealp{He2017}), the asynchronized variations between the absorption line systems and the quasar continuum reveal that the outflow clouds we studied in J1208+0355 are in relatively high ionization states with ionization parameters of Log$U$ larger than --1.8.}

The last factor is that the higher ionization ions, such as N\,{\footnotesize V} and C\,{\footnotesize IV} ions, are much stronger in their absorption strengths than the lower ionization lines like Si\,{\footnotesize IV} (Table \ref{tab.1} and Figure \ref{fig.2}), indicating that these outflow clouds are in a relatively high ionization state. This situation of a high ionization state is consistent with the inference from the asynchronized variations between the absorption lines and those of the quasar emission continuum. 

Based on the analyses above , we ascribe the absorption line changes {(at least for system B}) in J1208+0355 to the IC scenario, which is a response to the changes in the quasar’s continuum flux. 

\subsection{{Line-locking Signature within system B}} \label{sec.3.2} 
Line-locking is an observational signature that the successive shielding of the outflow clouds locks the clouds themselves together in outflow velocity. This observational signature is usually interpreted as an evidence for the radiative acceleration (e.g., \citealp{Foltz1987,Braun1989,Srianand2000,Srianand2002,Ganguly2003}). The line-locking signature in NALs has been generally identified from both BAL and non-BAL quasars (e.g., \citealp{Foltz1987,Srianand2002,Hamann2011,Bowler2014}). \citet{Lu2018complex2} presented the report of the line-locking phenomenon of blended NALs within trough-like BALs in quasar SDSS J021740.96--085447.9. In J1208+0355, we once again found the line-locking signature of the blended NALs within a BAL (system B; see Table \ref{tab.2}), which supports the idea that these outflow clouds are driven by the radiative force caused by resonance lines. In addition, the line-locking signatures in J1208+0355 indicate that the observer’s sight line may be almost parallel to the wind streamlines (e.g., \citealp{Hamann2011}). 

\subsection{{Kinematic Shift of system A}} \label{sec.3.3} 
As shown in Figure \ref{fig.3}, the kinematic shift of system A was clearly detected in the C\,{\footnotesize IV} ion. The kinematic shift of a BAL can be caused by several reasons, such as the following three situations: directional shift in the outflow, actual acceleration of the outflow clouds in our line of sight (LOS), and changes in velocity-dependent quantities, for example, the ionization state or covering factor and so on (e.g., \citealp{Hall2002,Hall2007,Gabel2003}). We hold the view that actual LOS acceleration driven by radiation pressure may be the mechanism responsible for the kinematic shift in J1208+0355. Our view is on the basis of several observational facts. {One weak suggestive argument is that radiative forces may} play an important role in driving the BAL outflow in J1208+0355, according to the line-locking phenomena shown in system B (Section \ref{sec.3.2}). Second, recombination-driven EW variations that respond to the continuum variability are confirmed for both systems A and B (Section \ref{sec.3.1}), indicating the effect from changing radiation energy as well as a close physical connection between these two systems. Both the above two factors reveal that the BAL outflow of J1208+0355 is affected by the background radiation energy. So we infer that the kinematic shift shown in system B may be produced by actual LOS acceleration of the outflow clouds, which is driven by radiation pressure from the background light source.

\section{{Conclusion}} \label{sec.4} 
We have analyzed two BAL systems (systems A and B) of quasar J1208+0355, based on its two-epoch spectra from the SDSS. The main discussions we have presented are as follows:

1.The EW variations of the absorption lines {(at least for system B)} in J1208+0355 are caused by the IC scenario in response to the variability of the ionization continuum according to the following observational factors: first, the two BAL systems, as well as multiple NALs in system B show coordinated strengthenings; second, the continuum flux shows an obvious weakening; third, the N\,{\footnotesize V} and C\,{\footnotesize IV} BALs are much stronger than the Si\,{\footnotesize IV}, which is in agreement with the conclusion of the high ionization state from the asynchronized variations between the absorption line systems and the ionization continuum.

2.The NALs within system B show line-locking phenomena, indicating that these outflow clouds are driven by the radiative force caused by resonance lines.

3.One of the BAL absorption line systems, system A, shows kinematic shift in its C\,{\footnotesize IV} ion. Because the above two research aspects convincingly reveal that the BAL outflows are affected by the background radiation energy, we draw a conclusion that the kinematic shift shown in system B may be produced by actual LOS acceleration of the outflow clouds, which is driven by radiation pressure from the background light source.

\acknowledgments
We acknowledge the anonymous referee for comments that helped to improve this paper. This research was supported by the Guangxi Natural Science Foundation (2017GXNSFAA198348) and the National Natural Science Foundation of China (11903002).

Funding for the Sloan Digital Sky Survey IV was
provided by the Alfred P. Sloan Foundation, the U.S.
Department of Energy Office of Science, and the Participating
Institutions. SDSS-IV acknowledges support and resources
from the Center for High-Performance Computing at the
University of Utah. The SDSS website is \url{http://www.sdss.org/}.

SDSS-IV is managed by the Astrophysical Research
Consortium for the Participating Institutions of the SDSS
Collaboration including the Brazilian Participation Group, the
Carnegie Institution for Science, Carnegie Mellon University,
the Chilean Participation Group, the French Participation
Group, Harvard-Smithsonian Center for Astrophysics, Instituto
de Astrofísica de Canarias, The Johns Hopkins University,
Kavli Institute for the Physics and Mathematics of the Universe
(IPMU)/University of Tokyo, Lawrence Berkeley National
Laboratory, Leibniz Institut für Astrophysik Potsdam (AIP),
Max-Planck-Institut für Astronomie (MPIA Heidelberg),
Max-Planck-Institut für Astrophysik (MPA Garching), MaxPlanck-Institut für Extraterrestrische Physik (MPE), National
Astronomical Observatories of China, New Mexico State
University, New York University, University of Notre Dame,
Observatário Nacional/MCTI, The Ohio State University,
Pennsylvania State University, Shanghai Astronomical Observatory, United Kingdom Participation Group, Universidad
Nacional Autónoma de México, University of Arizona,
University of Colorado Boulder, University of Oxford,
University of Portsmouth, University of Utah, University of
Virginia, University of Washington, University of Wisconsin,
Vanderbilt University, and Yale University.

\bibliographystyle{aasjournal}
\bibliography{NALvsBALandshift} 

\end{document}